\documentclass{PoS}

\usepackage{epsfig}
\usepackage{amsmath}
\usepackage{amssymb,amsfonts}
\usepackage{graphicx}
\usepackage{graphics}
\usepackage{amsthm}
\usepackage{graphicx}
\usepackage{multirow}
\usepackage{float}

\newcommand{\cA}{{\cal A}}
\newcommand{\cAb}{{\overline{\cal A}}}

\newcommand{\cQ}{{\cal Q}}
\newcommand{\cU}{{\cal U}}
\newcommand{\cN}{{\cal N}}

\newcommand{\vn}{ {\bf n} }

\newcommand{\hatbmu}{\widehat{\boldsymbol {\mu}}}

\def\bec{\begin{center}}
\def\eec{\end{center}}
\def\beq{\begin{equation}}
\def\eeq{\end{equation}}
\def\bea{\begin{eqnarray}}
\def\eea{\end{eqnarray}}

\title{Supersymmetric gauge theories on the lattice: Pfaffian phases and the Neuberger $0/0$ problem}

\ShortTitle{The sign problem and the Neuberger $0/0$ problem for lattice gauge theories}

\author{\speaker{Dhagash Mehta}\thanks{This work was supported by the U.S. Department of Energy grant under contract no. DE-FG02-85ER40237 and Science Foundation Ireland grant 08/RFP/PHY1462. Simulations were performed using USQCD resources at Fermilab. AJ's work is also supported in part by the LDRD program at the Los Alamos National Laboratory. We thank Martin Schaden for his useful comments.}\\
        Department of Physics, Syracuse University, Syracuse, New York 13244, USA.\\
        E-mail: \email{dbmehta@syr.edu}}

\author{Simon Catterall\\
        Department of Physics, Syracuse University, Syracuse, New York 13244, USA.\\
        E-mail: \email{smc@physics.syr.edu}}

\author{Richard Galvez\\
        Department of Physics, Syracuse University, Syracuse, New York 13244, USA.\\
        E-mail: \email{ragalvez@syr.edu}}

\author{Anosh Joseph\\
        Theoretical Division, Los Alamos National Laboratory, Los Alamos, New Mexico 87545, USA\\
        E-mail: \email{anosh@lanl.gov}}

\abstract{Recently a class of supersymmetric gauge theories have been successfully implemented on the lattice. However, there has been an ongoing debate on whether lattice versions of some of these theories suffer from a sign problem, with independent simulations for the ${\cal N} = (2, 2)$ supersymmetric Yang-Mills theories in two dimensions yielding seemingly contradictory results. Here, we address this issue from an interesting theoretical point of view. We conjecture that the sign problem observed in some of the simulations is related to the so called Neuberger $0/0$ problem, which arises in ordinary non-supersymmetric lattice gauge theories, and prevents the realization of Becchi-Rouet-Stora-Tyutin symmetry on the lattice. After discussing why we expect a sign problem in certain classes of supersymmetric lattice gauge theories far from the continuum limit, we argue that these problems can be evaded by use of a non-compact parametrization of the gauge link fields.}

\FullConference{The XXIX International Symposium on Lattice Field Theory - Lattice 2011\\
July 10-16, 2011\\
Lake Tahoe, California, USA}

\begin{document}
{\bf Introduction}: There exist at least two major reasons to study supersymmetric gauge theories regularized on a space-time lattice. Firstly, supersymmetric Yang--Mills (SYM) theories, an important subclass of these gauge theories, play prominant role in the AdS/CFT correspondence in which it is conjectured that string theories in AdS backgrounds can be described via a conformal supersymmetric gauge theory living on the boundary of the AdS space. At finite temperature this holographic duality relates properties of black holes in the AdS space to the thermal behavior of this holographic dual gauge theory. This duality is such that the supergravity limit of the string theory corresponds to the strongly coupled regime of the gauge theory. In such a situation one can hope to study the non-perturbative regime of the gauge theory using lattice methods - see \cite{Catterall:2007fp-Catterall:2008yz-Catterall:2009xn-Catterall:2010fx} for some recent lattice results. Secondly, supersymmetric field theories are among the most promising candidate models for the physics beyond the Standard Model but many questions related to supersymmetry breaking are inherently non-perturbative in character and call for study using strong coupling methods such as lattice simulations.

In spite of these obvious motivations, little progress has been made in studying supersymmetric theories on the lattice until relatively recently. The problem is that naive discretizations of supersymmetric theories break supersymmetry completely leading to a fine tuning problem to regain supersymmetry as the continuum limit is taken. However, recently a number of new approaches have been advocated that address this problem and new classes of supersymmetric lattice gauge theories constructed - see \cite{Kaplan:2002wv, Cohen:2003xe-Cohen:2003qw, Catterall:2004np, Catterall:2005fd, Catterall:2009it, arXiv:1110.5983} and \cite{Sugino:2003yb, Sugino:2004qd, D'Adda:2005zk-D'Adda:2007ax-Kanamori:2008bk-Hanada:2009hq-Hanada:2010kt-Hanada:2010gs-Hanada:2011qx}. In this paper we focus on just one of these new lattice formulations in which a part of the algebra of (extended) supersymmetry is maintained under discretization. This approach can be used to directly formulate a lattice theory with ${\cal N}=4$ supersymmetry in four dimensions and can be derived in two seemingly different ways. The first proceeds via orbifold projection of a matrix model \cite{Kaplan:2002wv} while the second approach employs twisting of the spinorial supersymmetries of the target (continuum) SYM theory \cite{Sugino:2003yb, Sugino:2004qd, Catterall:2004np}. However it has been shown that both yield equivalent lattice constructions \cite{Unsal:2006qp, Takimi:2007nn, Damgaard:2007xi, Catterall:2007kn, Catterall:2009it}. The lattice theories constructed this way are gauge invariant, preserve a subset of the original supersymmetries and provide the target (continuum) theories in  the limit of vanishing lattice spacing. These constructions are successful only in implementing certain SYM theories with extended supersymmetries that include an extensive list of interesting class of SYM theories such as $\cN=2$ in $2$D, $\cN=4$ in $3$D and $4$D, and their dimensional reductions.

However, even after implementing the SYM models on the lattice we potentially encounter an additional, and rather crucial, difficulty - the fermionic sign problem: Consider the lattice theory with a set of bosons $\phi$ and fermions ($\overline{\psi}$, $\psi$). The partition function of the theory is
\bea
\label{eq:general_Z}
Z &=& \int [d\phi][d\overline{\psi}][d\psi] \exp\Big(-S_B[\phi] - \overline{\psi} M[\phi]\psi\Big)= \int [d\phi] Pf(M)~\exp \Big(-S_B[\phi]\Big)~,
\eea
where $M$ is the fermion matrix which is an anti-symmetric matrix and $Pf(M)$ is the Pfaffian of $M$. For a $2n \times 2n$ matrix $M$, the Pfaffian is explicitly given as $Pf(M)^{2} = \mbox{Det}\, M$. The Pfaffian, being a complex quantity, has a phase angle $\alpha$ that may change for different configurations. The $\alpha$-weighted average then may cancel between different configurations resulting in a huge statistical error. This is the famous (fermionic) sign problem.

In the lattice supersymmetry constructions, there has been an ongoing debate on the existence of the sign problem in the two-dimensional four supercharge ($\cQ=4$) lattice SYM theory. The resolution of this sign problem is very crucial as the extraction of continuum physics from this lattice model depends very much on it. In \cite{Giedt:2003ve}, it was shown that there did exist a sign problem in the lattice $\cQ=4$, $D=2$ SYM. However, since those results were obtained without any importance sampling of the partition function they would not necessarily hold in the full theory. However, in \cite{Catterall:2008dv} it was shown that for the phase quenched ensemble at non zero lattice spacing there still appeared to be a sign problem for this theory. Later, Hanada et al. \cite{Hanada:2010qg} showed that there was no sign problem for this theory in the continuum limit. In this work, we argue that for this theory, one must expect a sign problem in this model due to an extension of the so called Neuberger $0/0$ problem if one employs the usual (group based) parametrization of the gauge fields. However, if one employs a different non-compact parametrization of the bosonic fields, such as those used in \cite{Hanada:2010qg} it may be avoided. In the following, we first explain why the usual gauge-fixing  Faddeev-Popov (FP) procedure fails in lattice field theory giving rise to the conventional Neuberger $0/0$ problem. Then we show that the partition function for the $\cQ=4$ SYM theories, being $\cQ$-exact, can be viewed as being obtained through a Faddeev-Popov gauge-fixing procedure and hence potentially subject to the Neuberger $0/0$ problem, the latter appearing as a fermionic sign problem. However, we argue that the usual Neuberger problem only occurs if the relevant
group manifold contains non-trivial cycles - a feature which is only true if an exponential parametrization of the gauge links is made; a non-compact parameterization side steps the issue.

{\bf The Faddeev-Popov procedure and the Neuberger $0/0$ problem on the lattice}: Let us consider a gauge theory on a flat Euclidean space-time lattice defined by the partition function $Z = \int [D U] \exp(-S[U])$, where the gauge fields $U_a(\vn)$, defined along the links $(\vn, \vn + \hatbmu_a)$, with $\hatbmu_a$ being the basis vectors on the lattice, take values from a group $G$. Since the action $S[U]$ possesses a gauge symmetry $U_a(\vn) \rightarrow g(\vn)U_a(\vn)g^\dagger(\vn+\hatbmu_a)$, there are infinitely many degrees of freedom for the gauge configurations and hence the partition function is ill-defined. Here, all the gauge transforms $g(\vn)$ also take values from $G$. To remove the redundant degrees of freedom, one has to choose a representative out of the infinitely many possible physically equivalent configurations for a given set of gauge fields $U_a(\vn)$. This can be achieved by imposing a constraint on the gauge fields, called a gauge-fixing condition. In the perturbative limit, there is a well-defined procedure, the FP procedure, to impose the gauge-fixing condition in the path integral: to impose a condition $f(U, g)=0$, one can insert a unity defined by $Z_{GF} := 1 = \int D g \delta(f(U,g)) \Delta_{FP}$, and then integrate out the gauge transforms. Here, $\Delta_{FP}$ is the Jacobian determinant of $f(U,g)$ with respect to the $g$ variables, and $Z_{GF}$ is called the gauge-fixing device. Writing $\Delta_{FP}$ in terms of Grassmann variables (ghost and antighost fields), we obtain a gauge-fixed action which possesses a symmetry among gauge fields and ghosts and anti-ghosts, called the Becchi-Rouet-Stora-Tyutin (BRST) symmetry. Letting $\cQ$ be the BRST operator, the gauge-fixed part of the action can be written as $S_{GF} = \cQ \Lambda$, where $\Lambda$ is a combination of terms involving gauge fields, ghosts and anti-ghosts. This feature of $S_{GF}$ is called $\cQ$-exactness (and $S_{GF}$ is called a $\cQ$-exact term).

A crucial assumption here is that $f(U,g)=0$ should have a unique solution. While in the perturbative limit, because $f(U,g)=0$ turns out to be a linear equation, there is a unique solution for such a condition, Gribov showed that in the non-perturbative regime for any covariant gauge-fixing condition there is in general more than one solution, the additional solutions are called Gribov copies \cite{Gribov:1977wm}.

On the lattice with standard, i.e., group based Wilson parametrization of the gauge fields, there is another problem: the Gribov copies come in pairs with opposite signs of $\Delta_{FP}$. Thus the gauge-fixing partition function when summed over all Gribov copies is zero: $Z_{GF}=0$! Then, the expectation value of any gauge-fixed observable turns out to be $0/0$; this is known as the Neuberger $0/0$ problem \cite{Neuberger:1986vv}. This problem has a very important consequence: a BRST symmetry can not be realized on the lattice for the standard Wilson formulation. This problem was interpreted by Schaden in terms of a Witten-type topological field theory \cite{Witten:1988ze} and he arrived at the conclusion that on the lattice $Z_{GF}$
computes a topological invariant, the Euler characteristic, of the group manifold at each lattice-site, which is zero for $SU(N)$ or $U(N)$ \cite{Schaden:1998hz}.
(Also see \cite{arXiv:0710.2410, arXiv:0912.0450} for a recent point of view and \cite{Baulieu:1996rp} for the continuum counterpart.)

{\bf The lattice SYM partition function as a gauge-fixing partition function}: The 2D $\cQ=4$ SYM theory is also a $\cQ$-exact theory, i.e., the complete action of this theory can be written as $\cQ$ acting on some combination of bosons and fermions. In fact, this is a characteristic of a Witten-type topological field theory \cite{Witten:1988ze}. In this case the operator $\cQ$ (the nilpotent BRST charge of the topological field theory) arises via twisting of the continuum supersymmetries. In other words, we can construct the complete two-dimensional $\cQ=4$ SYM theory described in the twist and orbifold prescriptions via gauge-fixing the fields $(\cA_a, \cAb_a)$ to flat complexified
connections. We start with a trivial classical action $S_{\textrm{classical}}[\cA_a,\cAb_a]=0$, which exhibits a topological shift symmetry $\cA_a \rightarrow \cA_a' = \cA_a + \theta_a$ with $\theta_a$ being an infinitesimal (complexified) shift. The topological symmetry is then gauge-fixed by following the Batalin-Vilkovisky \cite{Batalin:1984jr} procedure. This process will generate a set of ghost, ghost-for-ghost and anti-ghost fields as end products, providing the necessary bosonic and fermionic terms in the action. We then relax the restrictions on the nilpotent BRST charge $\cQ$ and the topological nature of correlation functions to enhance the topological field theory to a supersymmetric gauge theory. The final action we achieve through this process is the supersymmetric lattice action. Thus the corresponding partition function of the twisted supesymmetric lattice theory
can be viewed as the corresponding $Z_{GF}$ in the FP procedure for the classical action $S_{\textrm{classical}} = 0$. Another important feature of the Witten-type topological field theories is the existence of a Nicolai map: after integrating out the fermionic fields in the partition function (i.e., the one given in Eq.~(\ref{eq:general_Z})), for the Witten-type topological field theories, there exist a change of variables (i.e., a map) that trivializes the bosonic action $S_B[\phi]$ and the Jacobian of that change of variables cancels the Pfaffian up to its sign. The final expression of $Z$ can be shown to compute a topological invariant, namely the Euler characteristic (Since in the twisted SUSY theories, the gauge-fields are complexified, the actual topological invariant the partition function computes may not be precisely the Euler characteristic but some closely related topological quantity.). This means that this theory, when put on the lattice, should suffer from (an extended version of) the Neuberger $0/0$ problem. The difference here is that the Gribov copies now correspond to vacuum states of the supersymmetric lattice theory. The Neuberger $0/0$ problem rephrased {\it a la} Schaden would imply that these vacuum states come in pairs with opposite signs of the Pfaffian and cancel each other - a sign problem.

{\bf Differences between the two contemporary simulations approaches}: The preceding argument assumes that the lattice gauge fields are represented as exponentials of the continuum fields. Rather remarkably, these supersymmetric lattice constructions allow for another possibility; a non-compact parameterization based on expanding the fields on the algebra of the corresponding $U(N)$ group. In the first approach the complexified gauge fields appearing in the twisted continuum theory $\cA_a(x) = A_a(x) + iB_a(x)$, are mapped to link fields $\cU_a(\vn)$ living on the link $(\vn, \vn+\hatbmu_a)$ through the mapping $\cU_a(\vn) = e^{\cA_a(\vn)}$. The gauge links belong to the group $G$, which can be taken as the complexified $U(N)$ group. We call this parametrization the {\it exponential parametrization}. However, since the fields live on complexified $U(N)$ another possibility is open to us; namely we can simply expand the fields on the {\it algebra} of $U(N)$ and require that the theory generate a vacuum expectation value for the imaginary part of the trace mode of the link fields. That is, $\cU_a(\vn) = {\mathbf I} + \cA_a(\vn)$. In  order to not break $U(N)$ gauge invariance one must then supplement the action with a gauge invariant potential which {\it automatically} generates the required vev for the trace mode (such a vacuum expectation value is compatible with the moduli space of the theory). Since this expansion agrees with the usual expansion of the exponential in the Wilson formulation the two parametrizations lead to the same naive continuum limit. However, they can clearly be very different at non-zero lattice spacing. The exponential representation was used in \cite{Catterall:2008dv}, while in \cite{Hanada:2010qg} the non-compact representation was used.

We have added a potential term of the form \cite{Hanada:2010qg}: $S_M = \mu^2 \sum_\vn [(\frac{1}{N}){\rm Tr}(\cU_a^\dagger(\vn) \cU_a(\vn))-1]^2$ to the lattice action, with $\mu$ a tunable mass parameter, which can be used to control the expectation values and fluctuations of the lattice fields. Notice that such a potential obviously breaks supersymmetry -- however because of the exact supersymmetry at $\mu = 0$ all supersymmetry breaking counterterms induced via quantum effects will possess couplings that vanish as $\mu \to 0$ and so can be removed by sending $\mu\to 0$ at the end of the calculation.

We conjecture that the sign problem for the two-dimensional twisted SYM with exponential parametrization is nothing but an extended Neuberger $0/0$ problem. For the non-compact parametrization we live only in the algebra and hence the topology of the group manifold is irrelevant. Interestingly, independent of the lattice SYM scenario, non-compact gauge fields have also recently been used to restore BRST symmetry on the lattice \cite{arXiv:0710.2410}. There the decompactification of the gauge fields is done via stereographic projection of the original compact gauge-fields. It should be noted that the Neuberger $0/0$ problem applies strictly only to the lattice theory and it is not clear how it behaves as we approach the continuum limit. Note that the Neuberger $0/0$ problem is a sufficient condition to have a sign problem in the corresponding theory and not a necessary condition, i.e., there may still be sign problem in theories where the extended Neuberger $0/0$ problem does not apply. In the case of two- and four-dimensional $\cQ=16$ twisted SYM, the action has a $\cQ$-closed piece in addition to a $\cQ$-exact piece so Neuberger will {\it not} apply independent of the parametrization of the gauge links. Of course, this theory may still have a sign problem originating from some other source although recent results in fact indicate that this theory is safe from a sign problem for both zero and non-zero lattice spacing \cite{Catterall:2011aa}. Below we provide some results from the recent numerical simulations to support our conjecture.

{\bf Simulation results}: To simulate the theory, we rescale all the lattice fields by powers of the lattice spacing to make them dimensionless. This leads to an overall dimensionless coupling parameter of the form $N/(2\lambda a^2)$, where $a=\beta/T$ is the lattice spacing, $\beta$ is the physical extent of the lattice in the Euclidean time direction and $T$ is the number of lattice sites in the time direction. The lattice coupling is defined as $\kappa = (NT^2)/(2\lambda\beta^2)$, for the symmetric two-dimensional lattice, i.e., the spatial length $L=T$. Note that $\lambda$ is the dimensionless physical `t Hooft coupling. In our simulations, the continuum limit can be approached by fixing $\lambda\beta^{2}$ with $N$ fixed and increasing the lattice size $L \rightarrow \infty$. We take our parameters to be $\lambda = 1.0, 2.0$ and $L = 2, \cdots, 12$.

We examined the Pfaffian phase measured in the phase quenched ensemble generated over sequential runs of about 20000 Monte Carlo configuration (MCC) time steps for each $L$ and $\lambda$. Measurements were taken after every $5$ MC time steps. The simulations we show were performed for anti-periodic (thermal) boundary conditions for the fermions, and for both non-compact and exponential parameterizations of the gauge links.
\begin{figure}[t]
\bec \vspace{-0.7cm}
\includegraphics[width=8.0cm, height=3.8cm]{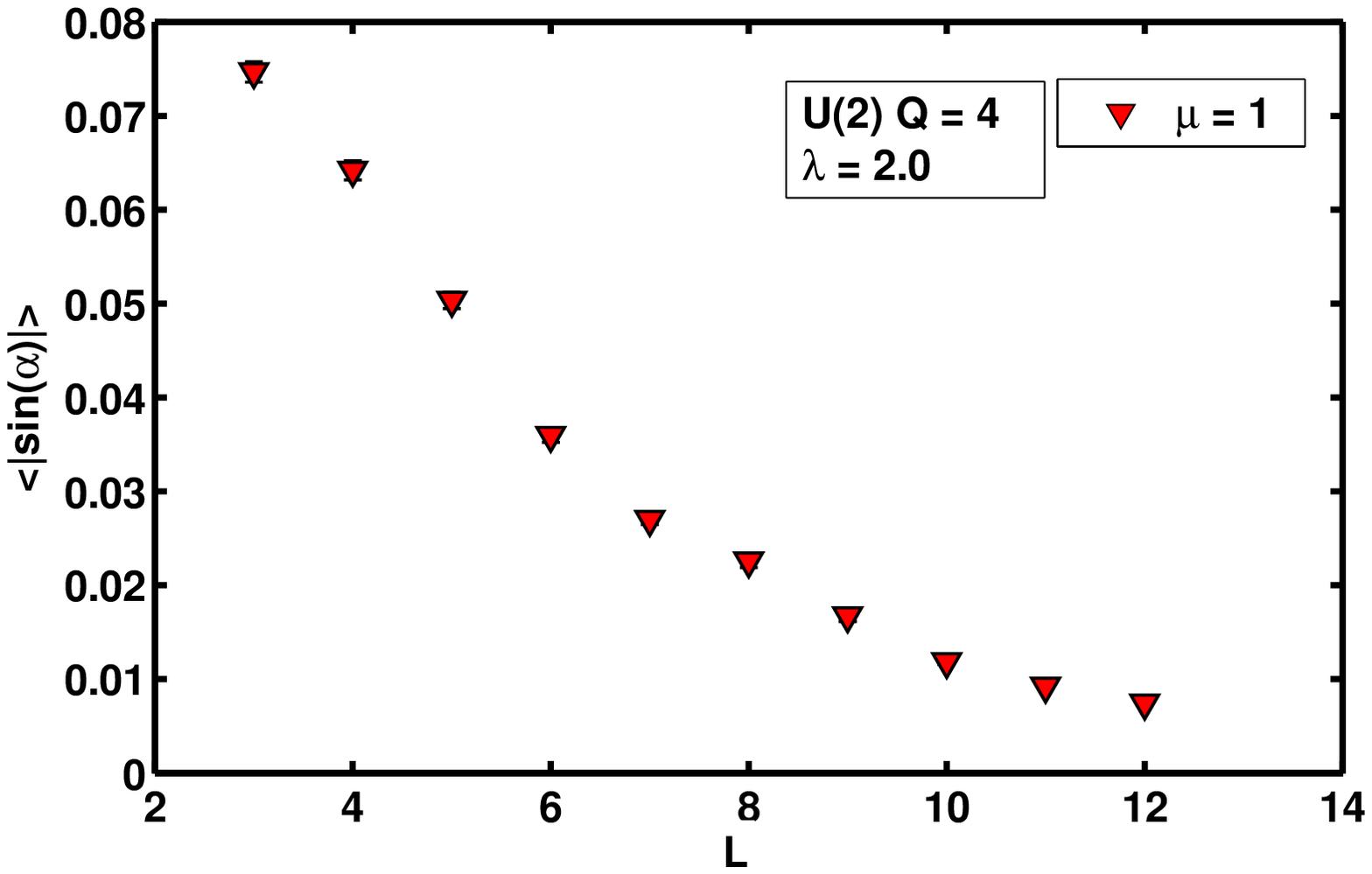}\includegraphics[width=6.0cm, height=3.8cm]{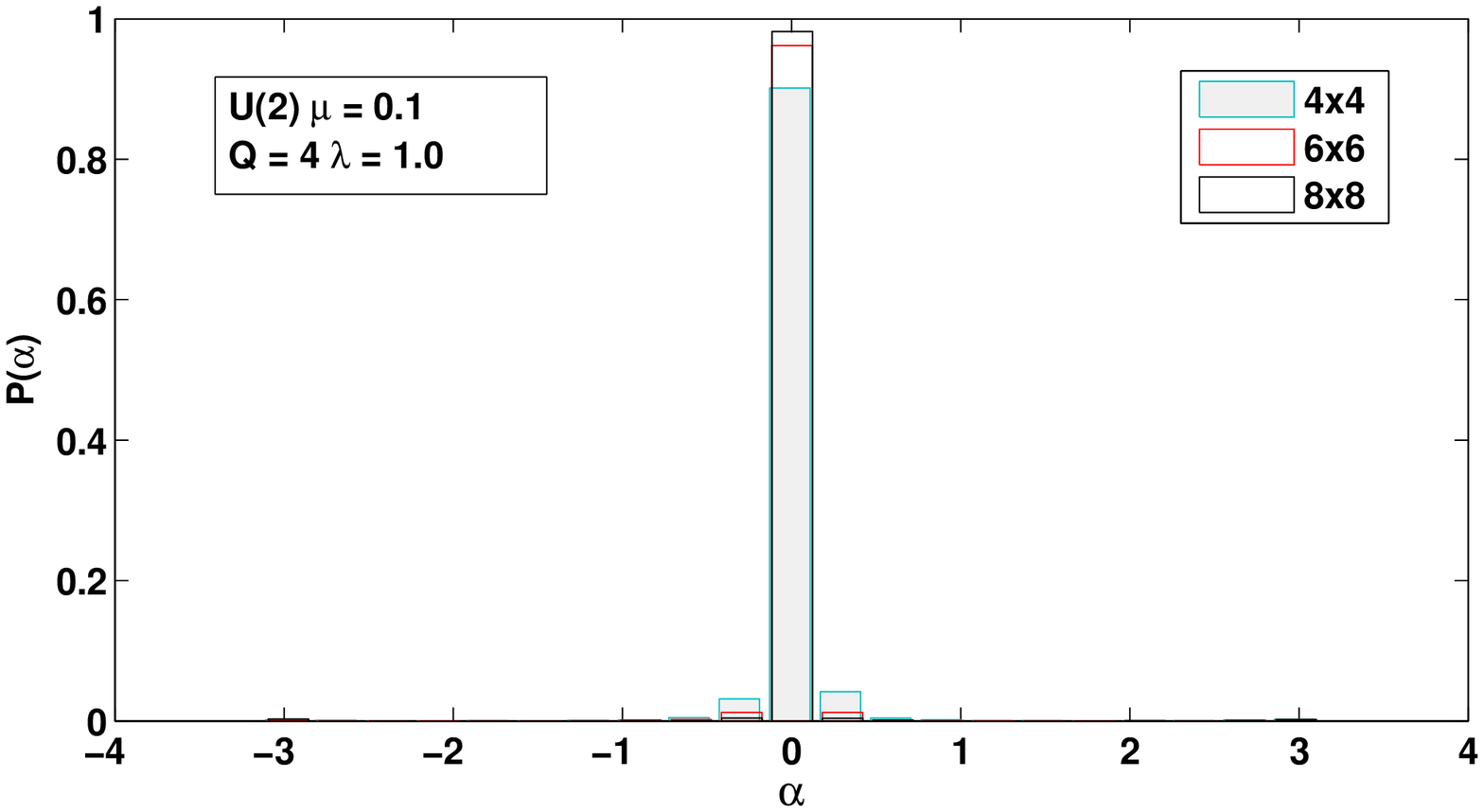}
\vspace{-0.75cm}
\caption{\label{fig:U2_Q4_2D_linear}The histograms of the phase angles $\alpha$ for $\cQ =4$ theories with gauge group $U(2)$, $\lambda = 1.0$, $\mu=0.1$, for non-compact representation.}
\eec
\end{figure}
\begin{figure}[t]
\bec \vspace{-0.7cm}
\includegraphics[width=6.4cm, height=3.8cm]{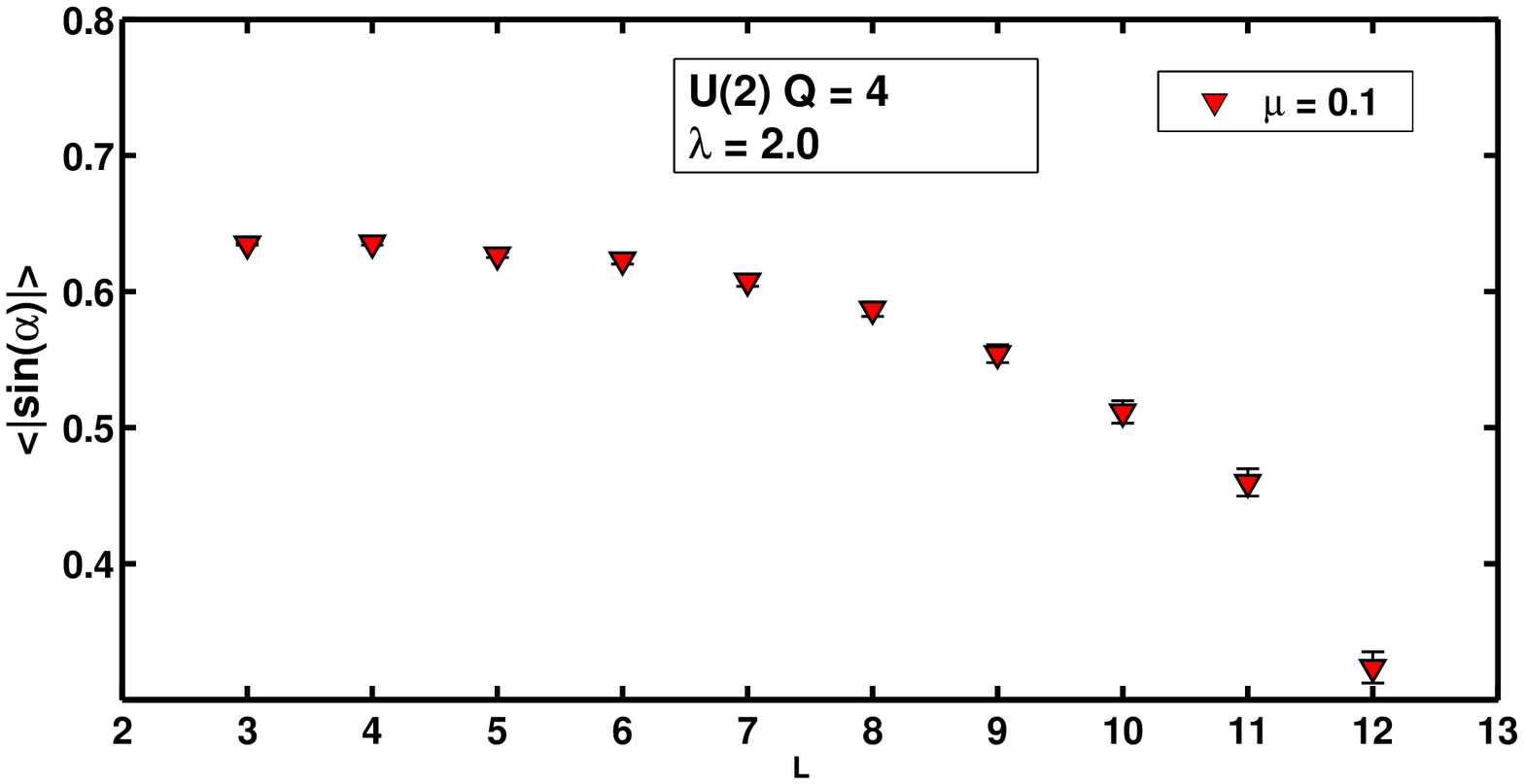}~~~~~~~\includegraphics[width=6.4cm, height=3.8cm]{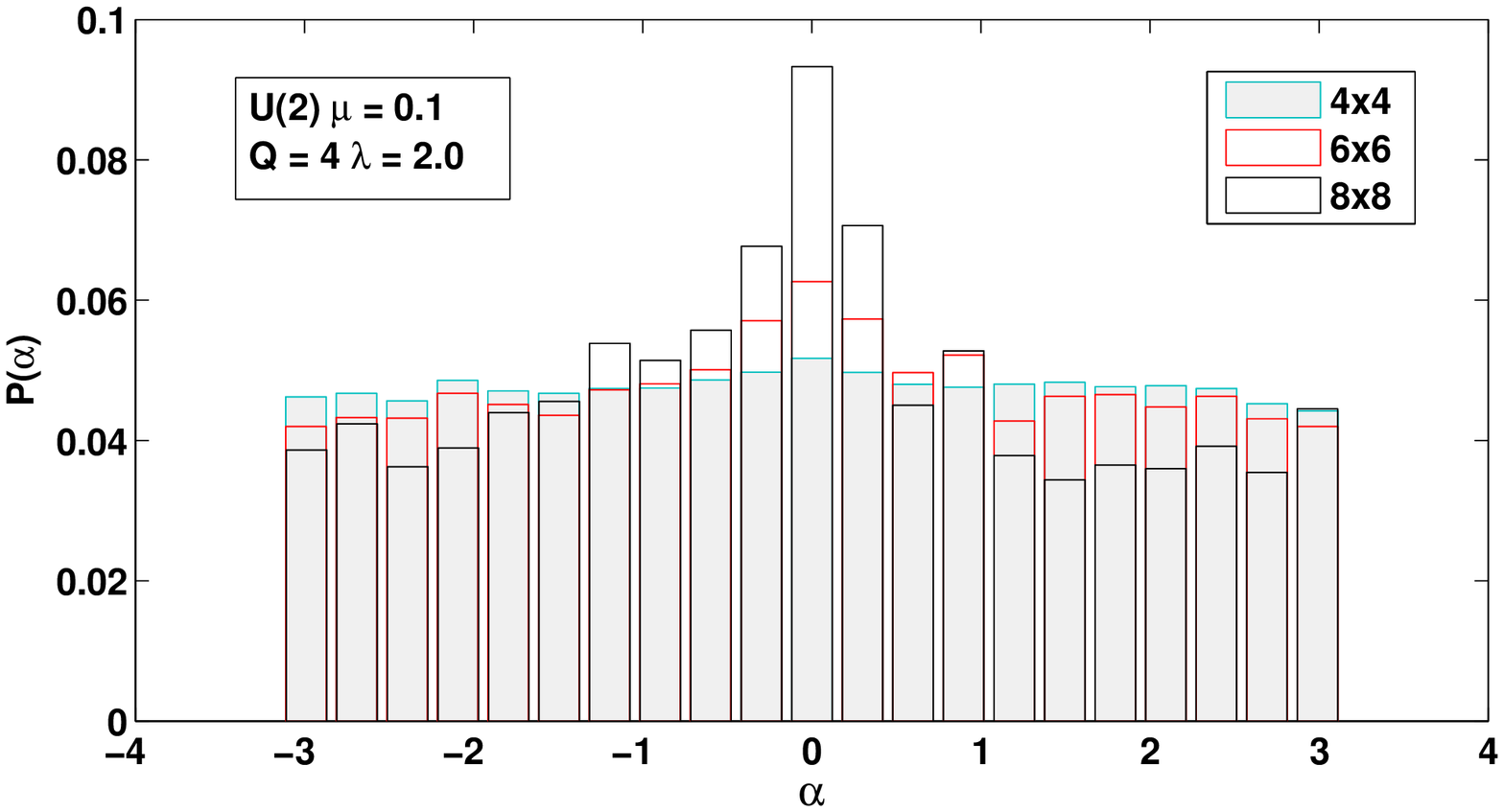}
\vspace{-0.4cm}
\caption{\label{fig:U2_Q4_2D_exponential}The histograms of the phase angles $\alpha$ for $\cQ =4$ theories with gauge group $U(2)$, $\lambda = 2.0$, $\mu=0.1$, for exponential representation.}
\eec \vspace{-0.75cm}
\end{figure}
The first plot shows the expectation value of the absolute value of $\sin{\alpha}$ ($\alpha$ is the Pfaffian phase) for the $\cQ=4$ model at $\mu=0.1$ using non-compact links together with a histogram of the phase angle. Clearly no sign problem is visible even on very coarse lattices. The second plot shows the same quantities using exponential links and the sign problem is visibly severe (in fact at its worst) away from the continuum in this case.

{\bf Conclusions}: In this work, we have discussed the possibility of a sign problem in two-dimensional lattice SUSY with twisted supersymmetries. It appears that the severity of any sign problem is heavily dependent on the parametrization of the gauge fields. For the non-compact parametrization, we show that there is no sign problem even on the smallest lattices while for the exponential parameterization the theory suffers from a severe sign problem away from the continuum limit. We have provided a possible theoretical explanation of this observed sign problem: with exponential parameterization, it is a manifestation of  the Neuberger $0/0$ problem.

\end{document}